\documentclass[12pt]{iopart}

\newcommand{\beq}{\begin{eqnarray}}
\newcommand{\eeq}{\end{eqnarray}}

\usepackage{graphicx}

\begin{document}

\title[Eccentricity and elliptic flow in $pp$ collisions at the LHC]{Eccentricity and elliptic flow in $pp$ collisions at the LHC}

\author{E~Avsar$^a$, Y~Hatta$^b$, C~Flensburg$^c$, J -Y~Ollitrault$^d$ and T~Ueda$^e$}
\address{${}^a$ 104 Davey Lab, Penn State University, University Park, 16802 PA, USA\\
 ${}^b$ Graduate School of Pure and Applied Sciences, University of Tsukuba,
Tsukuba, Ibaraki 305-8571, Japan\\
${}^c$ Dept. of Theoretical Physics, S\"olvegatan 14 A, S223 62 Lund, Sweden\\
${}^d$ CNRS, URA2306, IPhT, Institut de physique th\'eorique, CEA Saclay, F-91191
Gif-sur-Yvette, France \\
${}^e$ Institut f\"ur Theoretische Teilchenphysik,
Karlsruhe Institute of Technology (KIT),
D-76128 Karlsruhe, Germany}
\ead{
%eavsar@phys.psu.edu,
hatta@het.ph.tsukuba.ac.jp
%,christoffer.flensburg@thep.lu.se,
%jean-yves.ollitrault@cea.fr,ueda@particle.uni-karlsruhe.de
}
\begin{abstract}
High-multiplicity proton--proton collisions at the LHC may exhibit collective phenomena  such as elliptic flow. We study this issue using DIPSY, a brand--new Monte Carlo event generator which features almost--NLO BFKL dynamics and describes the transverse shape of the proton including all fluctuations. We predict the eccentricity of the collision as a function of the multiplicity and estimate the magnitude of elliptic flow. We suggest that flow can be signaled by a sign change in the four--particle azimuthal correlation.

\end{abstract}

%\maketitle

The observation of high--multiplicity events in $pp$ collisions  at the LHC opens up an interesting possibility that the collective flow, usually discussed in the context of nucleus collisions, may be realized in the final state of  $pp$ collisions \cite{Collaboration:2010gv}. Indeed,  highest multiplicity events in the 7 TeV $pp$ run have $dN_{ch}/d\eta \sim 40$ which is comparable to   semi--central Cu--Cu collisions at RHIC, and flow  has been observed in the latter. In this contribution we study some key questions about the possibility to observe elliptic flow \cite{Ollitrault:1992bk} in $pp$
 using DIPSY---a recently released Monte Carlo event generator \cite{Flensburg:2011kk}.  More details can be found in \cite{Avsar:2010rf}. For related works, see, \cite{Luzum:2009sb,
CasalderreySolana:2009uk,Pierog:2010wa}.

Firstly, one should bear in mind that high--multiplicity $pp$ collisions and nucleus collisions with the same multiplicity are vastly  different. In particular, the former are  {\it rare fluctuations} in the broad $N_{ch}$ distribution while the latter refer to average events at fixed centrality.  There are several sources of multiplicity fluctuations in $pp$: (i) Impact parameter fluctuation---Unlike in nucleus collisions, in $pp$ collisions it is not possible to determine the impact parameter for each event.
High multiplicity events mostly come from collisions with small impact parameter. (ii) Intrinsic fluctuation of the proton's wavefunction---Protons at the LHC undergo the QCD evolution to become a dense system of small--$x$ gluons.  Since the QCD evolution is stochastic, there are large event--by--event fluctuations in the gluon number. High multiplicity events arise from protons with an unusually large occupation number of gluons.   (iii) Fluctuation in the collision process---In a single high--multiplicity $pp$ event, there are many (more than 10) gluon--gluon scatterings. The number of subcollisions fluctuates due to the probabilistic nature of collisions  (partonic cross section).
(iv) Fluctuation in the final state parton showering---High multiplicity events typically contain several jets, and the fragmentation of jets is a stochastic process.

At first, point (i) seems to be a fatal blow to any hope of observing elliptic flow $v_2$ in $pp$. Naively, in central collisions the eccentricity would be very small, hence small, unobservable $v_2$.  However, this may be solved by point (ii). The QCD evolution generates fluctuations not only in the gluon number, but also in the transverse distribution of gluons because the gluon splitting probability depends on the transverse coordinates.\footnote{Previous works considered the fluctuation of `hot spots' \cite{CasalderreySolana:2009uk} and `flux tubes' \cite{Pierog:2010wa}. There the  transverse distribution of these objects was assumed to be random. In our case the transverse distribution of gluons is not random, but  governed by the QCD evolution. }  This makes it possible to have a sizable {\it participant eccentricity} even in central collisions
\beq
\epsilon_{\rm part} \equiv \frac{\sqrt{(\sigma_y^2-\sigma_x^2)^2+4\sigma^2_{xy}}}
{\sigma_y^2+\sigma_x^2}\,,
\label{epspart}
\eeq
where $\sigma_{x,y}$ are the variances of $x$ and $y$ coordinates of liberated gluons.  [By `liberated  gluons' we mean gluons which actually interacted as well as those in the underlying events.]
 We have computed (\ref{epspart}) using DIPSY \cite{Flensburg:2011kk,Avsar:2006jy}, a new event generator  which takes into account all the points (i)--(iv) above. It is based on the QCD dipole model \cite{Mueller:1993rr} which is the coordinate space formulation of the BFKL evolution, and therefore captures the correct transverse dynamics of small--$x$ gluons. In addition to the leading--order BFKL, DIPSY features a dominant part of the next--to--leading corrections, energy conservation, and saturation effects. Thanks to this, the energy dependence of observables is a prediction in DIPSY. Once the parameters have been fixed at one value of energy, there is no ad hoc re-tuning of parameters at different energies.

 %Fig.~\ref{dist} shows the multiplicity distribution at 7 TeV obtained in the current tune of DIPSY used in this work.
% \begin{figure}[ht]
% \begin{center}
%\includegraphics[width=80mm]{dist.eps}
%\end{center}
%\caption{
%Predictions for the eccentricity and the interaction area $S$ (left) and elliptic flow (right) versus %the  charged multiplicity in the interval $|\eta|<0.9$ at $\sqrt{s}=7$~TeV.}
%\label{dist}
%\end{figure}

\begin{figure}[ht]
 \begin{tabular}{c}
\includegraphics[width=50mm,angle=-90]{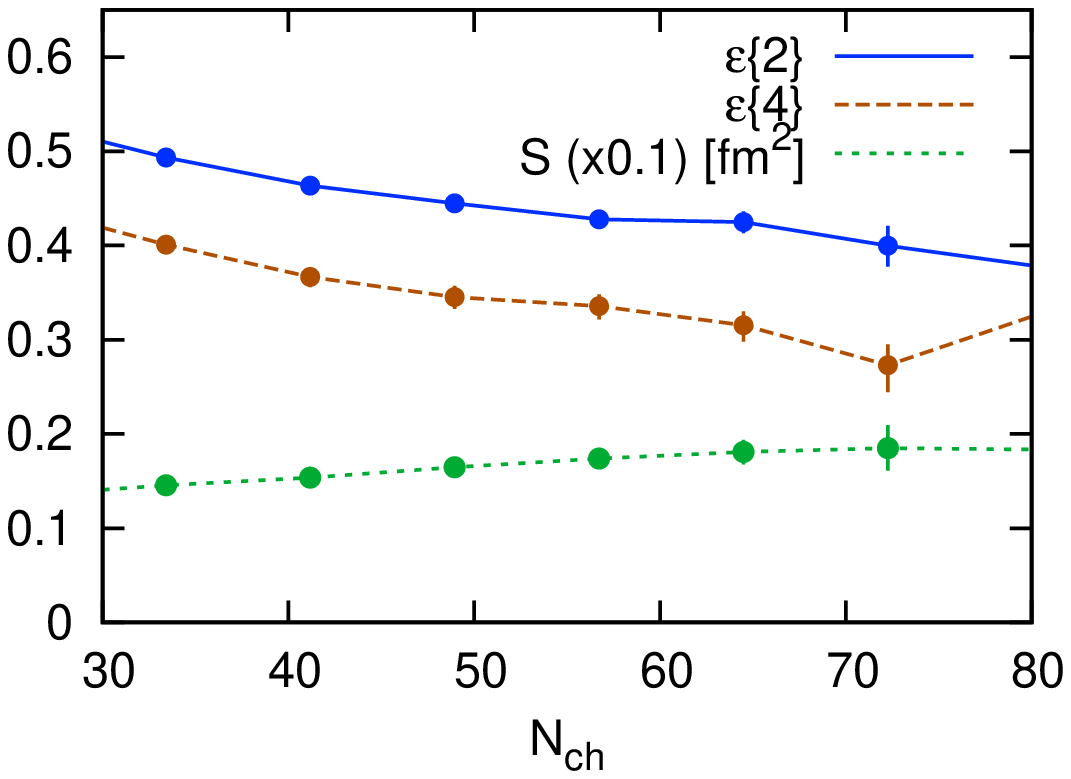}
\end{tabular}
\begin{tabular}{c}
\includegraphics[width=50mm,angle=-90]{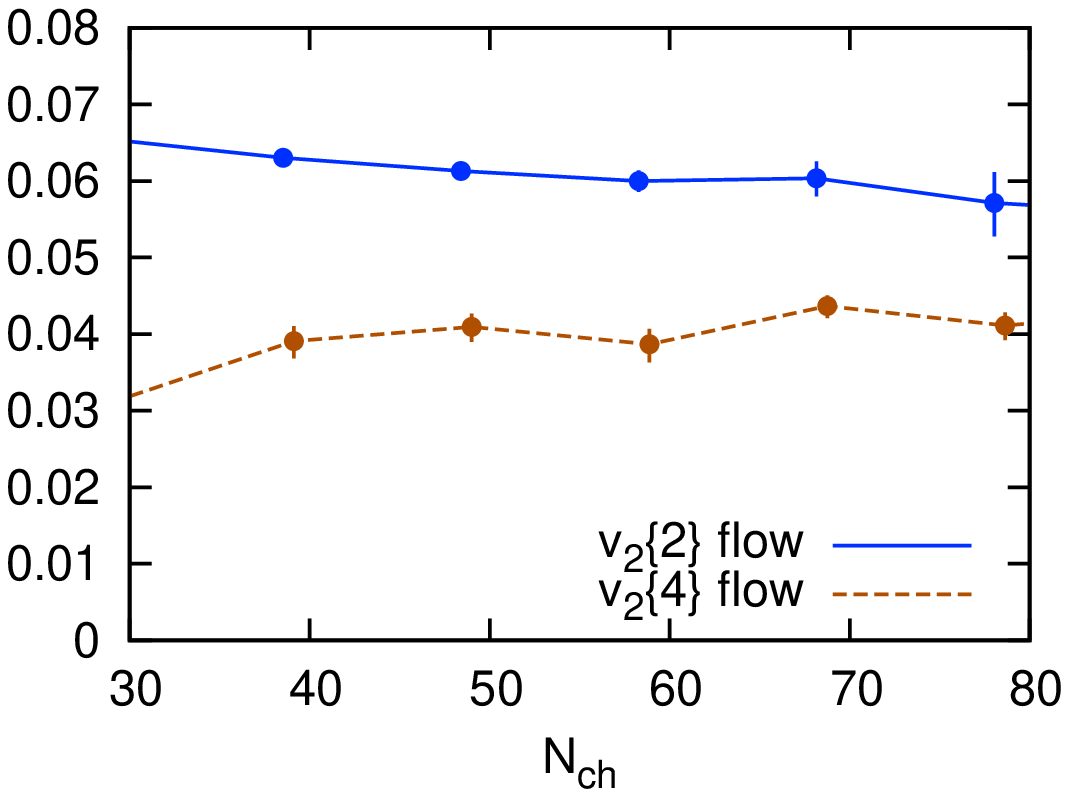}
\end{tabular}
\caption{
Predictions  for the eccentricity and the interaction area $S\,(\times 0.1) [{\rm fm}^2] $ (left) and elliptic flow (right) versus the  charged multiplicity in the interval $|\eta|<0.9$ at $\sqrt{s}=7$~TeV. }
\label{fig1}
\end{figure}

 The result for the eccentricity $\epsilon\{2\} \equiv \sqrt{\langle \epsilon_{\rm part}^2\rangle}$ and $\epsilon\{4\} \equiv ( 2 \langle \epsilon_{\rm part}^2 \rangle^2 - \langle \epsilon_{\rm part}^4 \rangle
)^{1/4}$ as well as the interaction area $\langle S\rangle$ at 7 TeV is plotted in Fig.~\ref{fig1}  (left) as a function of  $N_{ch}$ within the ALICE acceptance
$|\eta|<0.9$. [$\langle...\rangle$ denotes event--by--event averaging in a given multiplicity bin.] We see that the eccentricity is about 30--40\% in the highest multiplicity region. This is similar to the value in semi-central nucleus collisions at RHIC and the LHC.

Next we give an estimate $v_2$. In nucleus collisions, $v_2$ and $\epsilon$ are roughly proportional
\beq
v_2\{2\} = \epsilon\{2\}  \left ( \frac{v_2}{\epsilon} \right )_{{\rm hydro}} \, \frac{1}{1+\frac{\lambda}{K_0}
\frac{\langle S \rangle }{ \left\langle \frac{dN}{d\eta}\right\rangle}}\,,
\label{hydroform}
\eeq
 where  $\lambda/K_0$ is a certain parameter fitted to experimental data \cite{Drescher:2007cd}.
This empirical formula works both at RHIC and at the LHC, and  both for Au--Au and Cu--Cu at RHIC although they differ in size by a factor of two. The latter supports the general argument that
the applicability of hydrodynamics is controlled by the dimensionless parameter $\alpha \equiv \frac{\lambda}{K_0}\frac{S}{dN/d\eta}$ rather than the system size, and the necessary condition $\alpha < 1$ is well satisfied in high--multiplicity $pp$ events. We thus expect that (\ref{hydroform}) gives a  reasonable estimate of $v_2$ even in $pp$ collisions.
%, and this condition is better satisfied in high--multiplicity events (see Fig.~\ref{fig1}).
The result for $v_2\{2\}$ and $v_2\{4\}$ are plotted in Fig.~\ref{fig1} (right). [In this calculation we actually used a slightly improved version of (\ref{hydroform}), see, \cite{Avsar:2010rf}.]  Elliptic flow is about 6\%, comparable to the value found at the LHC.

 Is it possible to observe the flow contribution $v_2\sim 6\%$?  Experimentally, $v_2\{2,4\}$ are measured from the azimuthal angle correlation
  \beq
v_2^2\{2\}&=& \bigl\langle \cos (2(\phi_i - \phi_j))
\bigr\rangle \,,  \label{v2def}\\
v_2^4\{4\}  &=&
 2(v_2\{2\})^4 - \langle \cos(2(\phi_i +\phi_j-\phi_k -
\phi_l))\rangle \,. \label{v4def}
\eeq
  They differ from the genuine $v_2$ by the so--called nonflow contribution $
(v_2\{n\})^n = v_2^n + \delta_n$
 where $\delta_n$ is the $n$--particle correlations not associated with flow. In nucleus--nucleus collisions, they are relatively innocuous because they scale with the multiplicity as $
 \delta_n \sim 1/N_{ch}^{n-1}$.
In $pp$ collisions, however, one expects significant nonflow contributions from various initial and final state effects, and the ALICE collaboration has indeed found a much slower decrease in $N_{ch}$ \cite{Bilandzic:2011zz}. This slow decrease is also observed in
Monte-Carlo simulations, meaning that  the isolation of the flow contribution $v_2^n$ could be difficult for small values of $n$.
In the highest multiplicity bins, $v_2\{2\} \approx 0.13$ \cite{Bilandzic:2011zz} which is twice as large as the flow contribution. This implies that the two--particle correlation is dominated by nonflow effects.

 \begin{figure}[h]
 \begin{center}
\includegraphics[width=60mm,angle=-90]{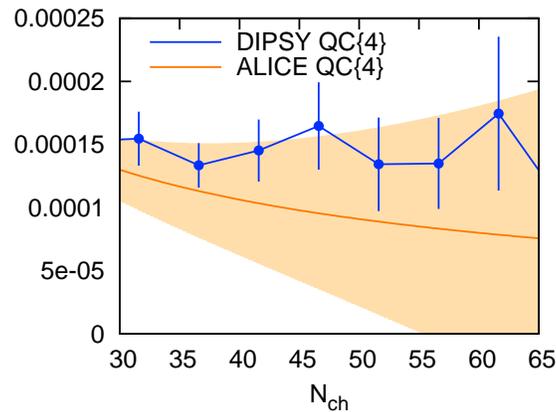}
\caption{
The DIPSY result for the fourth cumulant of the nonflow correlation $QC\{4\} \equiv -(v_2\{4\})^4$  compared with a rough parametrization of the preliminary ALICE data \cite{Bilandzic:2011zz}.}
\label{fig2}
\end{center}
\end{figure}

 We are thus led to turn to higher order cumulants $v_2\{n\}$ with $n\ge 4$ which are by definition insensitive to  two--particle nonflow correlations. Actually, the ALICE collaboration has also measured $v_2\{4\}$ \cite{Bilandzic:2011zz}. It turns out that the magnitude of $v_2\{4\}$ in the data is still larger than our flow prediction, but very interestingly,
  it has the `wrong' sign---the rhs of (\ref{v4def}) is negative!
  The same phenomena can be seen in PYTHIA and DIPSY (without flow effects), see Fig.~\ref{fig2}. On the other hand, in the flow scenario $(v_2\{4\})^4$ is positive. [Note, however, that the statistical error bars on the preliminary measured $(v_2\{4\})^4$ \cite{Bilandzic:2011zz}  are an order of magnitude larger than the value we predict from flow.]
  We thus propose to look for this sign change in experiment as a possible signature of flow:  If there is flow in the large $N_{ch}$ region, then the fourth order cumulant (\ref{v4def}),
which is negative in the absence of flow, will decrease in magnitude and eventually turn positive.

\end{document}